# Crossover of the high-energy spin fluctuations from collective triplons to localized magnetic excitations in doped $Sr_{14-x}Ca_xCu_{24}O_{41}$ cuprate ladders

Y. Tseng,[1,2,*] J. Thomas,[3] W. Zhang,[1] E. Paris,[1] P. Puphal,[4] R. Bag,[5] G. Deng,[6] T. C. Asmara,[1] V. N. Strocov,[1] S. Singh,[5] E. Pomjakushina,[4] U. Kumar,[3] A. Nocera,[7,8] H. M. Rønnow,[2] S. Johnston,[3,†] and T. Schmitt[1,‡]

[1]*Photon Science Division, Paul Scherrer Institut, CH-5232 Villigen PSI, Switzerland*
[2]*Laboratory for Quantum Magnetism, Institute of Physics, École Polytechnique Fédérale de Lausanne (EPFL), Switzerland*
[3]*Department of Physics and Astronomy, The University of Tennessee, Knoxville, TN 37996, USA*
[4]*Laboratory for Multiscale Materials Experiments, Paul Scherrer Institut, Villigen CH-5232, Switzerland*
[5]*Indian Institute of Science Education and Research, Dr. Homi Bhabha Road, Pune, Maharashtra 411008, India*
[6]*Australian Centre for Neutron Scattering, Australian Nuclear Science and Technology Organisation, New Illawarra Road, Lucas Heights, NSW 2234, Australia*
[7]*Stewart Blusson Quantum Matter Institute, University of British Columbia, 6224 Agricultural Road, Vancouver BC, Canada*
[8]*Department of Physics and Astronomy, University of British Columbia, Vancouver, British Columbia, Canada V6T 1Z1*

We studied the magnetic excitations in the quasi-one-dimensional (q-1D) ladder subsystem of $Sr_{14-x}Ca_xCu_{24}O_{41}$ (SCCO) using Cu $L_3$-edge resonant inelastic X-ray scattering (RIXS). By comparing momentum-resolved RIXS spectra with ($x = 12.2$) and without ($x = 0$) high Ca content, we track the evolution of the magnetic excitations from collective two-triplon (2T) excitations ($x = 0$) to weakly-dispersive gapped modes at an energy of 280 meV ($x = 12.2$). Density matrix renormalization group (DMRG) calculations of the RIXS response in the doped ladders suggest that the flat magnetic dispersion and damped excitation profile observed at $x = 12.2$ originates from enhanced hole localization. This interpretation is supported by polarization-dependent RIXS measurements, where we disentangle the spin-conserving $\Delta S = 0$ scattering from the predominant $\Delta S = 1$ spin-flip signal in the RIXS spectra. The results show that the low-energy weight in the $\Delta S = 0$ channel is depleted when Sr is replaced by Ca, consistent with a reduced carrier mobility. Our results demonstrate that off-ladder impurities can affect both the low-energy magnetic excitations and superconducting correlations in the $CuO_4$ plaquettes. Finally, our study characterizes the magnetic and charge fluctuations in the phase from which superconductivity emerges in SCCO at elevated pressures.

**Introduction**

Antiferromagnetic spin fluctuations are a promising candidate for mediating high-temperature superconductivity (HTSC) in cuprates [1]; however, the precise relationship between magnetism and superconductivity (SC) across the cuprate phase diagram remains unclear despite extensive studies. This uncertainty is due, in part, to the fact that modeling the competition between local moment formation and itinerant quasiparticles with doping is a tremendously difficult problem.

The quasi-one-dimensional (q-1D) spin ladders [Fig. 1] serve as an ideal platform to tackle this issue. Doped cuprate ladders share several similarities with two-dimensional (2D) cuprates, including high-energy spin-fluctuations [2], d-wave pairing [3–5], non-Fermi-liquid transport behavior [6], and large magnetic exchange couplings [7]. A SC phase has also been observed at elevated pressure ($1.5 - 8$ GPa) in the hybrid chain-ladder materials $Sr_{14-x}Ca_xCu_{24}O_{41}$ (SCCO) for $11 < x < 14$ [8–10]. Importantly, q-1D spin ladders are also more amenable to modeling [3–5]. Characterizing the magnetic excitations of doped cuprate ladders will, therefore, help us better understand HTSC in cuprates.

Magnetic excitations in even-leg ladders can be viewed as coupled spinons, i.e. triplons, due to the exchange interactions across their rungs [Fig. 1(b)] [3–5]. These gapped triplon excitations have been previously identified by experiments on undoped ladders [11,12], and are thought to be central to SC fluctuations upon doping [3–5]. Early studies also showed that the collective triplon excitations at high-energy closely resemble the prototypical "hourglass" magnetic excitations in 2D SC cuprates [13–15]. Conversely, experimental studies of triplons in doped ladders are sparse. Inelastic neutron scattering (INS) experiments discovered that the one-triplon (1T) gap at $q \equiv (q_{\text{leg}}, q_{\text{rung}}) = (\pi, \pi)$ survives up to $x = 12.2$, and is nearly unchanged compared to $x = 0$ in SCCO [16,17]. Nuclear magnetic resonance (NMR) studies on SCCO, on the other hand, find a decreasing triplon gap with increasing Ca content, which develops into a gap collapse at elevated pressure [18–20]. Although dispersing one- and two-triplon excitations have been reported in the nearly undoped ladder $Sr_{10}La_4Cu_{24}O_{41}$ using INS [12], comprehensive studies of their corresponding doping-dependence are lacking. Given that the spin fluctuations in 2D cuprates evolve in a non-uniform fashion across the Brillouin zone as a function of doping [1], a systematic assessment of the triplon response across a large portion of Brillouin zone, along with its relation to chemical pressure (Ca content), is crucial for understanding the role of spin fluctuations in the pairing in SCCO.

Owing to the strong spin-orbit coupling (SOC) in Cu $2p_{3/2}$ core-levels, Cu $L_3$-edge resonant inelastic X-ray scattering (RIXS) studies on cuprate materials can probe the momentum-dependent magnetic excitations in the dominant spin-flip channel ($\Delta S = 1$) [21]. Due to an order of magnitude difference in the magnetic exchange of the ladder (100 meV) and chain (10 meV) subsystems, a previous Cu $L_3$-edge RIXS study on $Sr_{14}Cu_{24}O_{41}$ showed that the high-energy magnetic excitations at around $100 - 350$ meV are dominated by the ladder triplons when probed at the Cu $3d^9$ resonance [22].

In this letter, we present a Cu $L_3$-edge RIXS study of the ladders in SCCO with varied Ca content ($x = 0$ and $x = 12.2$). Since superconductivity occurs in SCCO samples with $11 < x < 14$ under hydrostatic pressure [8–10], our measurements contrast the magnetic excitations of non-superconducting and superconducting samples. We observe a clear evolution in the low-energy RIXS spectra from collective two-triplon excitations to weakly-dispersive excitations as Sr is replaced by Ca. By comparing with numerical calculations of different disordered Hubbard ladder models, we find that the evolution of the spin excitations is consistent with hole-localization induced by off-ladder impurities. This conclusion is supported by a polarization analysis that uncovers spectral depletion of the charge-like excitations probed in the spin-conserving $\Delta S = 0$ scattering channel, reflective of suppressed carrier mobility. Our results naturally explain the observation of low-temperature insulating phases for high Ca content at ambient pressure, despite the expected increase of ladder holes [6,8,9]. Since SCCO serves as a bridge between q-1D and 2D cuprates, our study also helps elucidate the interplay between spin fluctuations, chemical pressure (i.e. Ca doping), electronic correlations, and SC in cuprate materials.

**Experimental Methods**

Single crystals of $Sr_{14}Cu_{24}O_{41}$ (Sr14) and $Sr_{1.8}Ca_{12.2}Cu_{24}O_{41}$ (Sr1.8Ca12.2) were grown with the traveling-solvent floating zone method [23,24]. The samples were cut and polished to form the a-c plane, with the b-axis pointing out-of-plane. Subsequently, top-post cleavage was performed *in-situ* before all measurements in a vacuum pressure of better than $5 \times 10^{-10}$ mbar. Cu $L_3$-edge RIXS and X-ray absorption spectroscopy (XAS) measurements were performed at the Advanced Resonant Spectroscopies (ADRESS) beamline at the Swiss Light Source (SLS), Paul Scherrer Institut [25–27]. The total energy resolution of the RIXS experiment was 95 meV at the Cu $L_3$-edge (~930 eV). The RIXS spectrometer was fixed at a scattering angle $2\theta = 130°$ while the experimental geometry was set such that the crystallographic b- and c-axis of the sample lay in the scattering plane [Fig. 1(a)]. This geometry allows no momentum-transfer along the ladder rungs and selectively enhances the multi-triplon RIXS response with even-parity along $q = (q_{\text{leg}}, q_{\text{rung}} = 0)$ [22,28]. RIXS measurements were acquired with 15 minutes per spectrum, and normalized to the total spectral weight of the crystal-field excitations. XAS spectra were recorded in total fluorescence yield (TFY) mode. $\sigma$ polarization was employed for the incident X-rays, unless specified otherwise. All measurements were taken at base temperature 20 K.

**Results and Discussions**

Figure 1(d) shows the Cu $L_3$-edge RIXS spectra at $q_{\text{leg}} = -0.31$ (rlu), where magnetic excitations appear around $200 - 400$ meV. These magnetic excitations have reduced spectral weight and increased broadening when Sr is replaced by Ca. Both RIXS spectra are taken at the $3d^9$ main resonance 930 eV indicated in Fig. 1(c). (The high-energy inelastic structures, including the local crystal-field splitting of the Cu $3d$ orbitals [dd excitations] from $-1.5$ to $-3$ eV loss, and the broad charge-transfer excitations around $-4$ eV loss and above, are shown in supplementary materials.)

Figures 2(a) and 2(d) display the momentum-dependent RIXS spectra along $q = (q_{\text{leg}}, 0)$. For Sr14 [Fig. 2(a)], we observe magnetic spectra from collective two-triplon excitations with a band minimum at the zone center, in accordance with the previous Cu $L_3$-edge work [22]. The excitations are broadened in Sr1.8Ca12.2 [Fig. 2(d)] and exhibit a flattened dispersion centered around 280 meV loss that extends across a large portion of the Brillouin zone.

To better understand our experimental results, we modeled the low-energy RIXS response to lowest order in the ultrashort core-hole lifetime expansion [29,30] using density matrix renormalization group (DMRG) calculations for doped Hubbard ladders [31,32]. This approximation has proven accurate for Cu L-edge measurements on strongly correlated ladders [42]. For Sr14, we adopted parameters resulting in magnetic exchange parameters close to the isotropic limit with $r = J_{\text{rung}}/J_{\text{leg}} = 0.85$ [33]. For Sr1.8Ca12.2, we estimated $r \approx 1.15$ for the highest Ca concentrations using perturbation theory and the available structural data (see supplementary materials). Using polarization-dependent O K-edge XAS measurements, we estimated the nominal ladder hole densities in Sr14 and Sr1.8Ca12.2. This procedure is facilitated by the atomic sensitivity of XAS to the distinct Cu-O bonding environments in the chain and ladder subsystems, enabling the evaluation of their hole content with O $2p$ character (see references in the supplementary materials). Using the model in Ref. [34], we obtain 0.06 and 0.11 (holes per Cu atom) for Sr14 and Sr1.8Ca12.2, respectively, in line with previous literature (see supplementary materials).

Figure 2 compares the experimental Cu $L_3$-edge RIXS data with DMRG calculations of the dynamical spin structure factor. Here, we consider both undoped and "nominally" doped ladders, as determined by our O K-edge XAS results. As in previous studies, the $\Delta S = 1$ scattering channel dominates the spectra [21]. The RIXS spectra for Sr14 are reasonably well described by both the undoped [Fig. 2(b)] and doped [Fig. 2(c)] models; however, we observe a slight increase in the broadening of the two-triplon excitations towards the zone boundary in the 6 % hole-doped ladders, which may be caused by the mixing with charge or Stoner-like continuum excitations like in 2D cuprates [35,36]. Overall, the experimental Sr14 Cu $L_3$-edge RIXS spectra are better described by the DMRG calculation of the 6 % nominally doped ladder.

The spin excitations for an 11 % clean hole-doped ladder in Fig. 2(e) are more diffuse compared to the undoped ladder calculations for Sr14 in Fig. 2(b), consistent with experiments. However, the increased hole doping also produces a pronounced downturn in the calculated dispersion close to the zone center, which does not occur in our experimental data for Sr1.8Ca12.2. Despite residual low-energy weight, the overall magnetic excitations for Sr1.8Ca12.2 [Fig. 2(d)] are clearly gapped and relatively dispersionless.

Since we have ruled out large values of the rung coupling, another mechanism must be responsible for the large change in magnetic excitations. Early transport studies found that the resistivity for SCCO with $x > 11$ decreased linearly down to $100 - 150$ K, followed by an abrupt upturn with further cooling [8,9]. The onset of this insulating behavior was attributed to carrier localization at low temperature [9]. We, therefore, consider the effects of localization on the low-energy excitations in Sr1.8Ca12.2. To this end, we performed DMRG calculations on ladders with additional impurity potentials distributed randomly throughout the ladder. The physical picture is that Ca doping introduces impurity potentials outside the ladder planes, which are poorly screened by the excess ladder charges. There are two possible scenarios here, which cannot be distinguished in our experiments. One is based on current SCCO literature, where it is widely accepted that Sr-Ca substitution transfers electrons from ladders to the edge-sharing $CuO_4$ chains [6,37] where they tend to localize [38] and exert an attractive potential on the ladder holes. (This scenario can be viewed as a sort of excitonic formation between the particle-hole pair introduced by the charge transfer process.) Alternatively, Ca could be acquiring excessive charges from the Sr sites, however, this is unlikely for two isovalent elements. Lastly, chemical doping in q-1D materials is often associated with structural distortions that lead to varied magnetic exchange couplings (see, for example Refs. [39,40]). We tested the possibility of the latter scenario and found it inconsistent with our estimated ladder-rung couplings in Sr1.8Ca12.2 (see supplementary materials).

To examine the feasibility of the localization proposal, we modeled the unscreened off-ladder impurities by introducing an extended impurity potential with 12.5 % coverage of ladder sites (see also the supplementary materials). The on-site value of the potential was set to $V_{\text{imp}} = 0.5U$ while the values on the neighboring sites up to next-nearest-neighbors was scaled as $\propto 1/r$, as sketched in the inset of Fig. 2(f). The resulting simulated RIXS signal in the $\Delta S = 1$ channel after averaging over eight disorder configurations are shown in Fig. 2(f). The downward dispersing spectral weight near zone center is suppressed, while the overall dispersion flattens close to the zone boundary. These modeled spectra are in closer agreement with the experiments and demonstrate that local impurities and charge localization are a plausible explanation for the flattened magnetic excitations. Our results thus suggest that other factors beyond simple charge doping like localization need to be invoked for explaining the evolution of the magnetic RIXS response with increasing Ca content.

To further confirm our interpretation of localized holes in the ladders of Sr1.8Ca12.2 we performed a polarimetric analysis of the RIXS measurements to assess the character of charge and spin scattering in the low-energy excitations [30,41–43]. Here, we apply the method established in Ref. [41] to disentangle $\Delta S = 1$ and $\Delta S = 0$ channels of the RIXS response, which was previously demonstrated for the two-spinon excitations of $CaCu_2O_3$ (see supplementary materials).

Figure 3 compares the $\Delta S = 1$ and $\Delta S = 0$ signal for $q_{\mathrm{leg}} = 0.38$ and 0.14 rlu. We find that the $\Delta S = 1$ channel indeed provides the major contribution to the low-energy RIXS response for both Sr14 and Sr1.8Ca12.2, confirming a magnetic origin for these excitations. In Sr14 [Fig. 3(a) and (b)], the $\Delta S = 0$ scattering is dominated by a sharp mode at slightly reduced energy compared to the $\Delta S = 1$ channel. Progress has recently been made in understanding the RIXS response in $\Delta S = 0$ channel of the ladder system, which indeed predicts a sharp bound two-triplon state in this channel [44]. Additionally, a broad component centered around $400 − 600$ meV appears in this channel, which is more pronounced near the zone boundary, coinciding with the momentum region where the excitation profiles show a stronger high-energy tail [Fig. 2(a)]. The energy scale of these excitations is comparable to the theoretically predicted charge and $\Delta S = 0$ multi-triplon excitations [10, 42]. Additionally, these modes resemble the 500 meV peak in the $\Delta S = 0$ RIXS channel of 2D cuprates [45–48], while its nature and connection to ladder excitations remains an open question.

The $\Delta S = 0$ channel in Sr1.8Ca12.2, on the other hand, is essentially a structure-less background below $-1$ eV loss, as shown in Fig. 3(c)-(d). Since these excitations arise from low-energy charge and $\Delta S = 0$ multi-triplon excitations, their suppression with Ca doping is consistent with localization. Previous theoretical studies on 2D cuprates and ladders demonstrated that the single spin-flip excitations dominate the Cu $L_3$-edge RIXS signal, whereas the $\Delta S = 0$ channel mainly consists of charge (particle-hole) excitations [30,33]. With mobile carriers, the $\Delta S = 0$ response is dominated by the charge continuum with excitations extending to zero-energy [30,33]. Conversely, with localized charges, the charge excitations are gapped with weight transferred to higher-energy [30,33]. This behavior is also captured by the calculated modified dynamical charge structure factor that reflects the RIXS intensity in this scattering channel (see supplementary materials). We, therefore, conclude that the observed non-dispersive and damped magnetic excitations in Sr1.8Ca12.2 can be understood using a hole-immobilized ladder picture, despite an increased carrier density.

Finally, we explore the implications of our results for superconductivity in the Ca-rich samples. Here, we envision that the localized carriers will be released as hydrostatic pressure is applied, eventually forming a superconducting condensate. Our calculated pair-pair correlation functions shown in Fig. 4 support this picture. Without impurities, we obtain a power-law dependence in the pair correlations, indicating strong pairing tendencies. The correlations are exponentially suppressed when we introduce the impurity potentials. Next, we mimic the effects of hydrostatic pressure by uniformly increasing the electronic hopping and find that the pair correlations begin to recover the robust behavior they had in absence of impurities. Despite the simplicity of our model (see Fig. 4 caption), these results imply that off-plane impurity potentials can significantly shape the low-energy magnetic and superconducting properties of the cuprate ladders. We believe that this issue should also be re-visited in 2D cuprates, where impurities in the charge reservoir layers are poorly screened.

Our results have important consequences for understanding the starting condition from which the SC state in SCCO is established with high Ca content and elevated pressures. The ingredients of a complete theory must now reconcile the roles of Sr-Ca substitution, magnetic and charge fluctuations, and the observed localization in Ca-doped SCCO. To further elucidate the overall effects of Sr-Ca replacement in SCCO, we suggest future experiments using INS at high-pressure or higher-resolution RIXS for measuring momentum-resolved magnetic excitations, which will help to clarify the underlying electronic correlations connected to pairing in SCCO.

**Conclusions**

We have performed Cu $L_3$-edge RIXS measurements on the spin ladder in SCCO. We observed a crossover in the magnetic excitation spectrum from the collective 2T excitations in Sr14 to a high-energy incoherent gapped magnetic mode in Sr1.8Ca12.2. By comparing with model DMRG calculations, we conclude that the observed reorganization of the magnetic excitations reflects a tendency towards carrier-immobility in the Ca-doped system. This conclusion was supported by polarization-dependent RIXS measurements, where we extracted a clear suppression of the $\Delta S = 0$ channel from well-defined charge and multi-triplon excitations ($x = 0$) to a featureless background ($x = 12.2$). We further calculated the pair correlations, and our results indicated that (Sr,Ca) layer impurities could impact the low-energy properties of the ladder subsystem. Our work calls for future studies re-examining the role of off-plane impurities in 2D cuprates in the doped regime where superconductivity, pseudogap, as well as charge and spin orders compete. For instance, INS experiments have revealed localized multi-magnon excitations in the $\Delta S = 1$ channel at ~130 meV, an energy scale consistent with the pseudogap values in 2D cuprates [49]. Finally, our results on localized excitations with Ca-doping could be useful to understand pair (pre-) formation in the pseudogap regime. On the other hand, our work gives valuable information on the dynamics of q-1D cuprate ladders, as well as their relationship with various competing electron-pairing hypotheses.

**Acknowledgements**

The experiments have been performed at the ADRESS beamline of the Swiss Light Source at the Paul Scherrer Institut (PSI). The work at PSI is supported by the Swiss National Science Foundation through project no. 200021_178867, and the Sinergia network Mott Physics Beyond the Heisenberg Model (MPBH) (SNSF Research Grants CRSII2_160765/1 and CRSII2_141962). T.C.A. acknowledges funding from the European Union's Horizon 2020 research and innovation programme under the Marie Skłodowska-Curie grant agreement No. 701647 (PSI-FELLOW-II-3i program). Y.T and T.S would like to thank V. Bisogni for valuable discussions. A. N. acknowledges, in part, funding from the Max Planck-UBC-UTokyo Center for Quantum Materials and the Canada First Research Excellence Fund, Quantum Materials and Future Technologies Program. J. T. and S. J. are supported by the National Science Foundation under Grant No. DMR-1842056. This work used computational resources supported by the University of Tennessee and Oak Ridge National Laboratory Joint Institute for Computational Sciences, and computational resources and services provided by Advanced Research Computing at the University of British Columbia.

*Present address: Department of Physics, Massachusetts Institute of Technology, Cambridge, Massachusetts 02139, USA; tsengy@mit.edu
†sjohn145@utk.edu
‡thorsten.schmitt@psi.ch

**Figures**

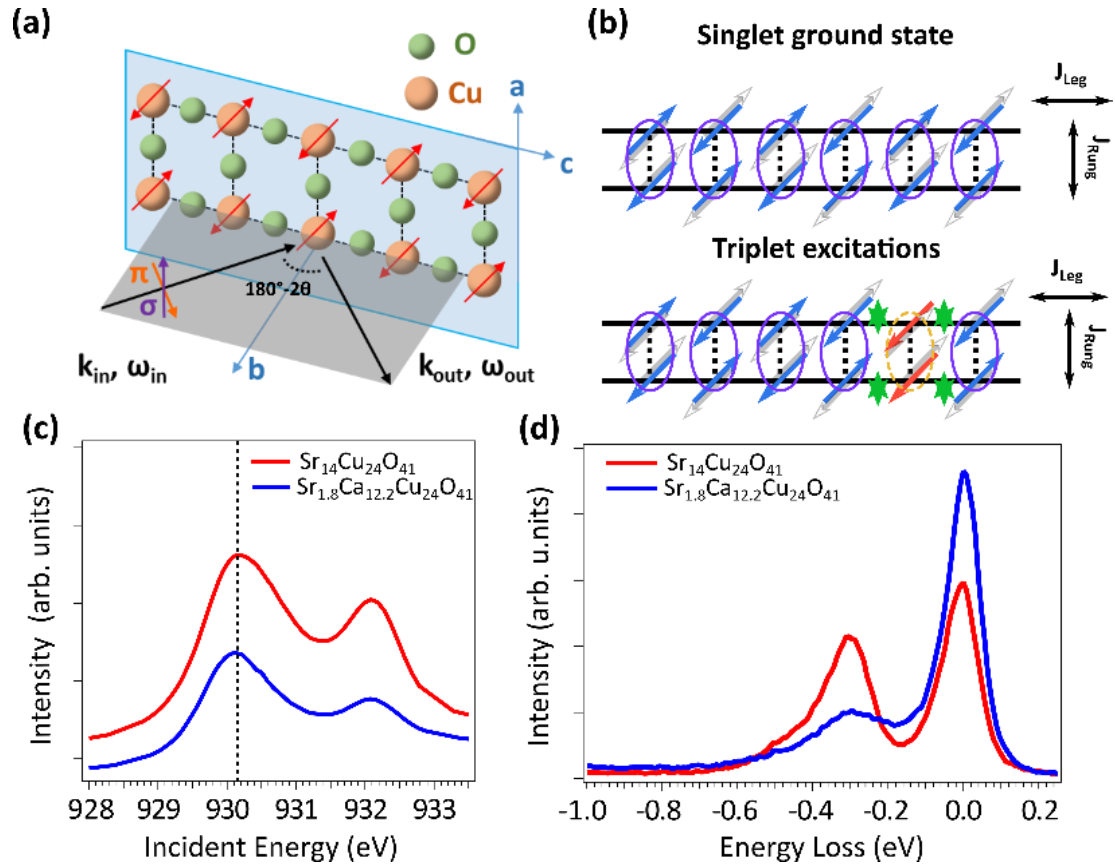

**FIG. 1** Schematics of (a) RIXS experimental geometry with the orientation of the two-leg $Cu_2O_3$ ladders in SCCO. The ladder-leg direction is along c-axis, with lattice constant $c_{leg}$ ~3.9Å. (b) $S = 1$ triplon excitations (orange dashed-line ellipse) in two-leg quantum spin ladders. The local antiferromagnetic interactions along the ladder leg and rung are mediated by spins (blue arrows) with equal probability pointing along any directions (opposite direction denoted as gray hollow arrows). The rung-singlets (purple ellipses) interact through ladder-leg exchange coupling. (c) Main resonances of Cu $L_3$-edge XAS. $3d^9$ white line is marked by the black dashed-line. (d) Cu $L_3$-edge RIXS spectra taken at $3d^9$ peak at $q_{leg} = -0.31$ (rlu) in $\sigma$ polarization. Here, a negative (positive) momentum transfer indicates the RIXS data taken in grazing incidence (emission) geometry.

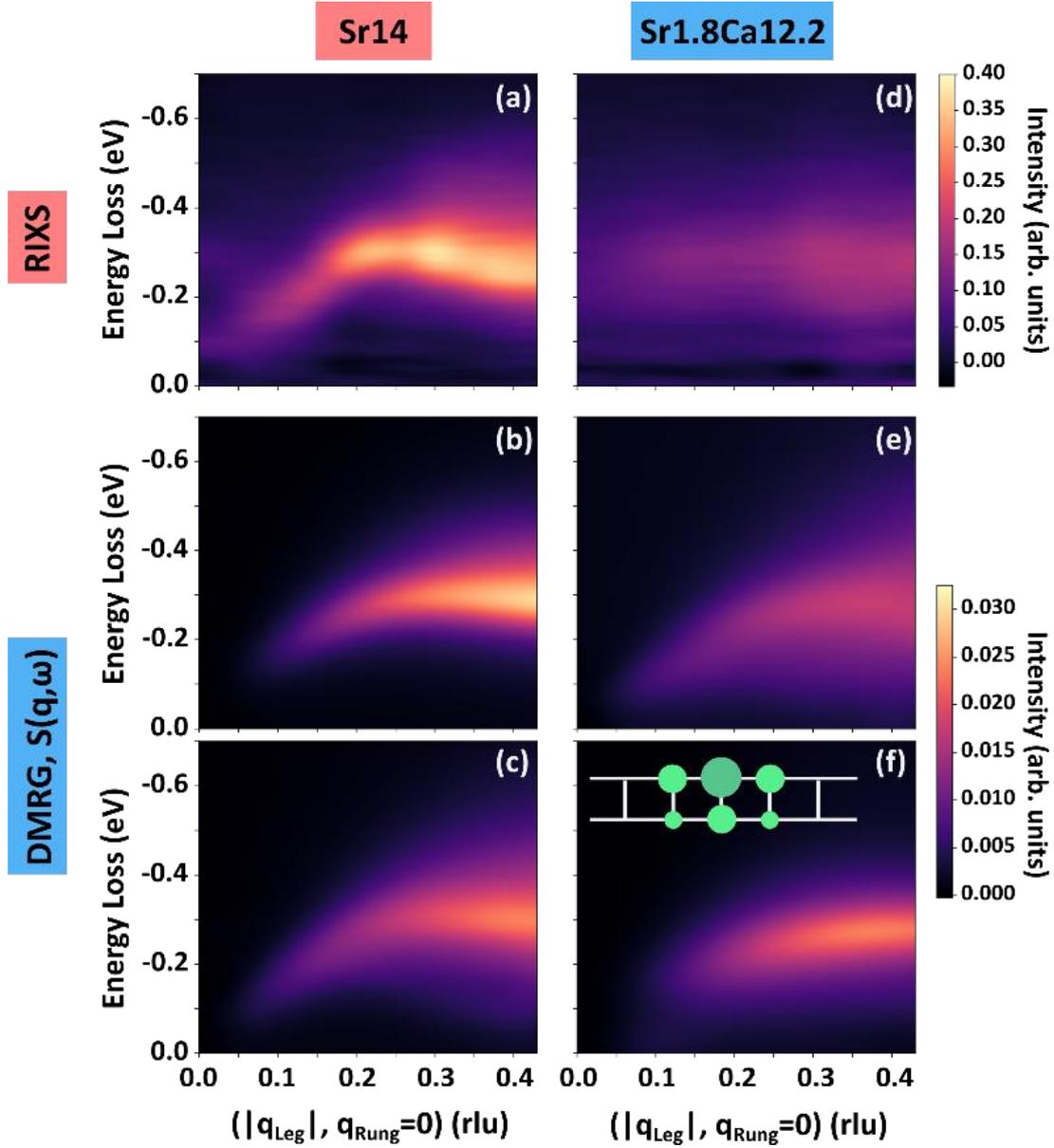

**FIG. 2** Left (right) panels show RIXS results for Sr14 (Sr1.8Ca12.2). The elastic line in each of the experimental spectra is subtracted for clarity. (a) and (d): Momentum-dependent RIXS experimental spectra for Sr14 and Sr1.8Ca12.2 in $\sigma$ polarization with bicubic interpolation. (b) Calculated $\Delta S = 1$ RIXS spectra on an undoped ladder cluster for Sr14. (c) and (e): Calculated $\Delta S = 1$ RIXS spectra on doped ladder clusters with hole doping p close to the experimentally determined values from O K-edge XAS results, which are about ~6 % and ~11 % for Sr14 and Sr1.8Ca12.2, respectively. The following set of parameters are applied for Sr14 (Sr1.8Ca12.2) on a $32 \times 2$ site cluster with spectral convergence: $t_{\text{leg}} = 340$ meV, $U = 8t_{\text{leg}}$, and $r = 0.85$ ($t_{\text{leg}} = 300$ meV, $U = 9t_{\text{leg}}$ and $r = 1.1457$). (f) Calculated $\Delta S = 1$ RIXS spectra for Sr1.8Ca12 on a $16 \times 2$ site doped ladder cluster with an impurity potential $V_{\text{imp}} = 0.5U$. We make additional assumptions that the on-site impurities (green circles) are screened poorly, and their potential extends to neighboring sites with strength (diameter size) scaled as $\propto 1/r$. The same $t_{\text{leg}}$, $U$ and $r$ parameters as in (e) are adapted in (f), with a hole doping of 12.5 %.

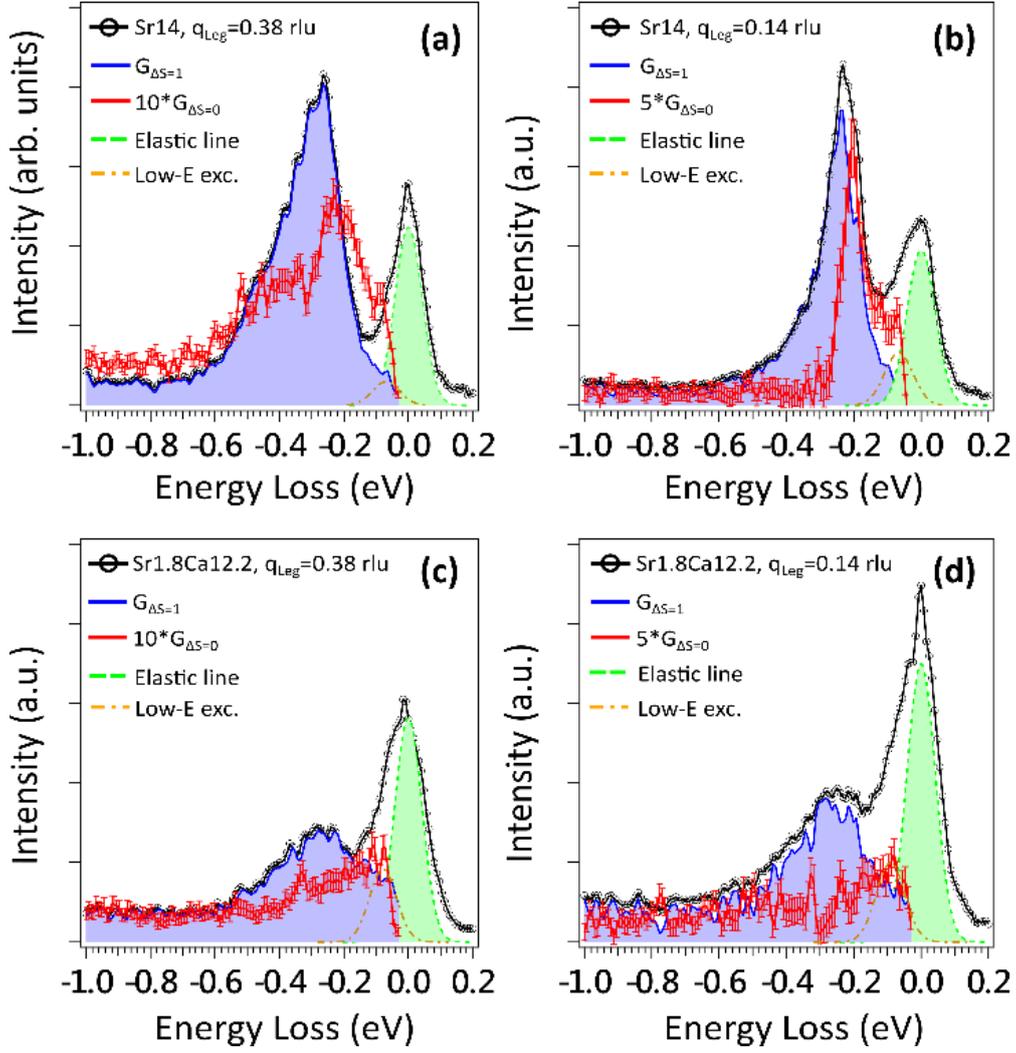

**FIG. 3** RIXS spectra with $\Delta S = 1$ (blue-shadowed) and $\Delta S = 0$ (red line) channels disentangled for (a)-(b) Sr14 and (c)-(d) Sr1.8Ca12.2 taken at $q_{\text{leg}} = 0.38$ rlu [(a) and (c)] and $q_{\text{leg}} = 0.14$ rlu [(b) and (d)]. Experimental data shown here are taken with π polarization in grazing emission geometry. We remind the reader that the RIXS intensity $I(\varepsilon, q)$ can be expressed as combination of $\Delta S = 1$ and $\Delta S = 0$ dynamical structure factors $G(\Delta S, \varepsilon, q, \omega)$ multiplied by form factors $F(\varepsilon, q)$ of the scattering processes [50–52].

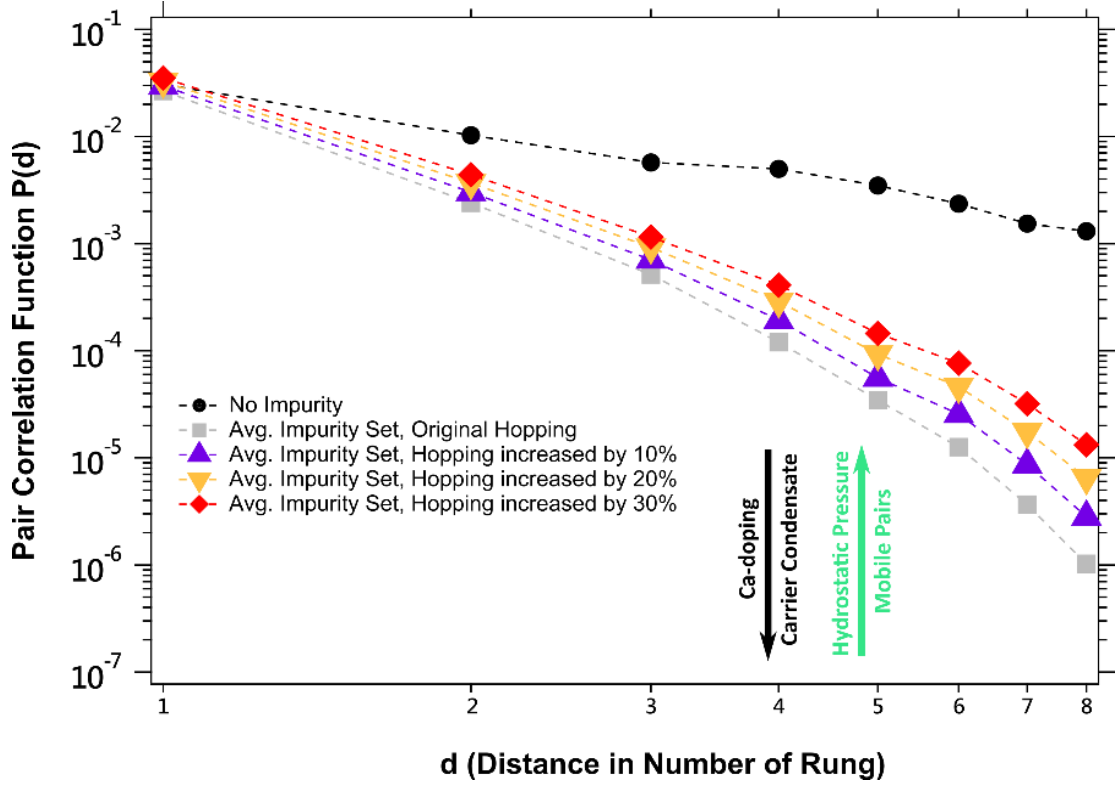

**FIG. 4** Rung-singlet pair correlations plotted as functions of distance along the ladder-leg direction in logarithmic scale (see methods in supplementary materials). For simplicity, the increasing hopping integral for simulating static pressure is uniform over the ladder cluster. The Hubbard repulsion and impurity potential $V_{\text{imp}}$ are kept fixed in these calculations.

# "Crossover of the high-energy spin fluctuations from collective triplons to localized magnetic excitations in doped Sr14-xCaxCu24O41 cuprate ladders" - Supplementary Material

Y. Tseng,[1,2,*] J. Thomas,[3] W. Zhang,[1] E. Paris,[1] P. Puphal,[4] R. Bag,[5] G. Deng,[6] T. C. Asmara,[1] V. N. Strocov,[1] S. Singh,[5] E. Pomjakushina,[4] U. Kumar,[3] A. Nocera,[7,8] H. M. Rønnow,[2] S. Johnston,[3,†] and T. Schmitt[1,‡]

[1]*Photon Science Division, Paul Scherrer Institut, CH-5232 Villigen PSI, Switzerland*
[2]*Laboratory for Quantum Magnetism, Institute of Physics, École Polytechnique Fédérale de Lausanne (EPFL), Switzerland*
[3]*Department of Physics and Astronomy, The University of Tennessee, Knoxville, TN 37996, USA*
[4]*Laboratory for Multiscale Materials Experiments, Paul Scherrer Institut, Villigen CH-5232, Switzerland*
[5]*Indian Institute of Science Education and Research, Dr. Homi Bhabha Road, Pune, Maharashtra 411008, India*
[6]*Australian Centre for Neutron Scattering, Australian Nuclear Science and Technology Organisation, New Illawarra Road, Lucas Heights, NSW 2234, Australia*
[7]*Stewart Blusson Quantum Matter Institute, University of British Columbia, 6224 Agricultural Road, Vancouver BC, Canada*
[8]*Department of Physics and Astronomy, University of British Columbia, Vancouver, British Columbia, Canada V6T 1Z1*

**Multi-peak Fitting of Magnetic Excitations in Cu $L_3$-edge RIXS**

Here, we show our peak assignment of the Cu $L_3$-edge RIXS response below $-1$ eV loss for $\mathrm{Sr_{14-x}Ca_xCu_{24}O_{41}}$ (SCCO) at selected momentum-transfer wave vectors. Fig. S1 shows RIXS spectra extending up to $-10$ eV loss for one momentum transfer vector along the leg direction. Fig. S2 shows the momentum-dependent RIXS measurements of SCCO with the incidence angle varied in the a-c plane from 10° to 65° in 5° steps. The fitted peak positions and widths are summarized in Fig. S3.

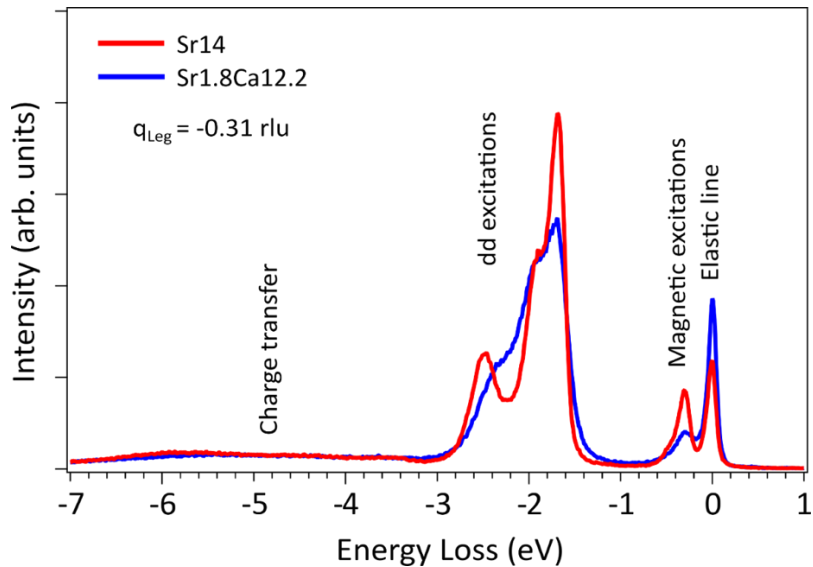

**FIG. S1:** Cu $L_3$-edge RIXS spectra with spectral components taken at $q_{\mathrm{leg}} = -0.31$ (rlu) for Sr14 and Sr1.8Ca12.2. From high- to low-energy we observe charge transfer excitations, inter-orbital crystal-field splitting of Cu 3d shell (dd-excitations), magnetic excitations and the elastic line.

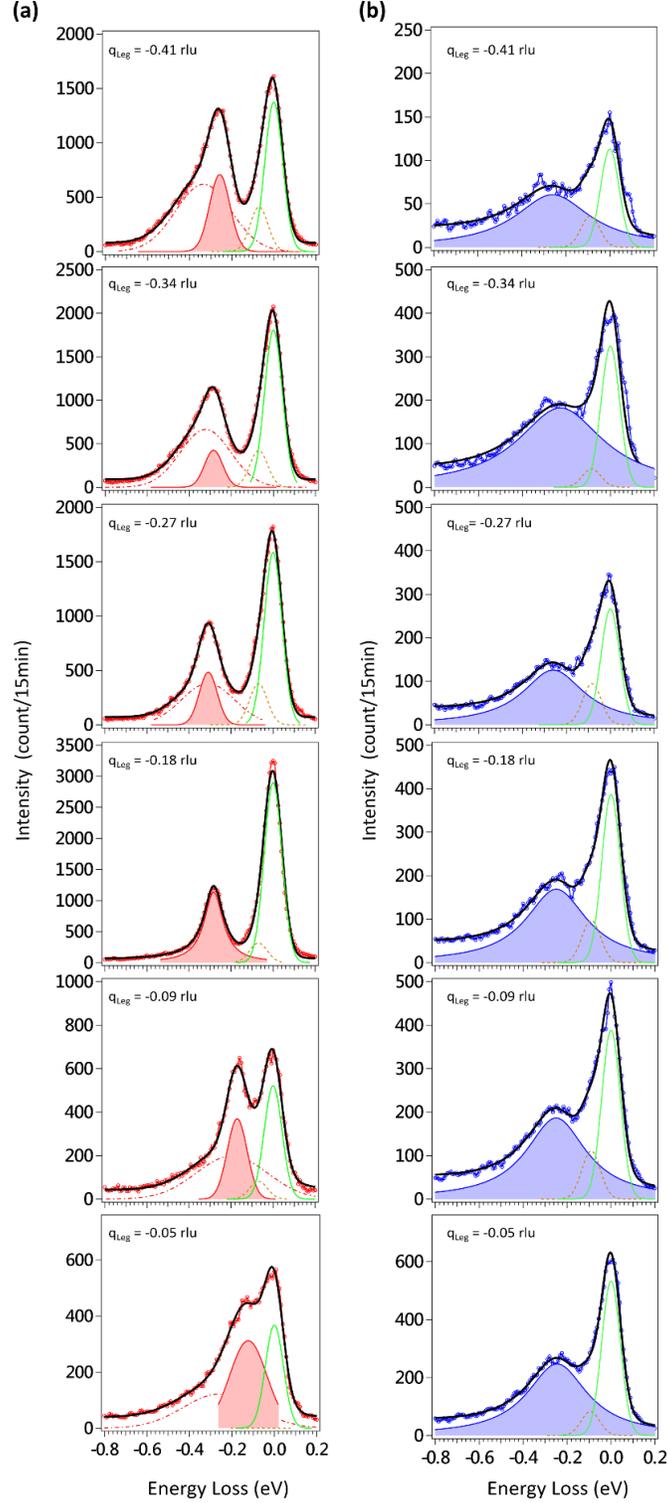

**FIG. S2:** Cu $L_3$-edge RIXS spectra with fitting components at selected momenta for (a) Sr14 and (b) Sr1.8Ca12.2. The energy loss peak around the 0 eV consists of the elastic (and diffuse) scattering (green solid lines) and quasi-elastic scattering (brown dash lines). Two-triplon (2T) excitations of Sr14 are fitted with two Gaussians, one lower-energy sharp mode (red solid line with shaded-region) and one broad high-energy component (red dashed line). 2T profiles in Sr1.8Ca12.2 are fitted by a damped Lorentzian (blue solid line with shaded region).

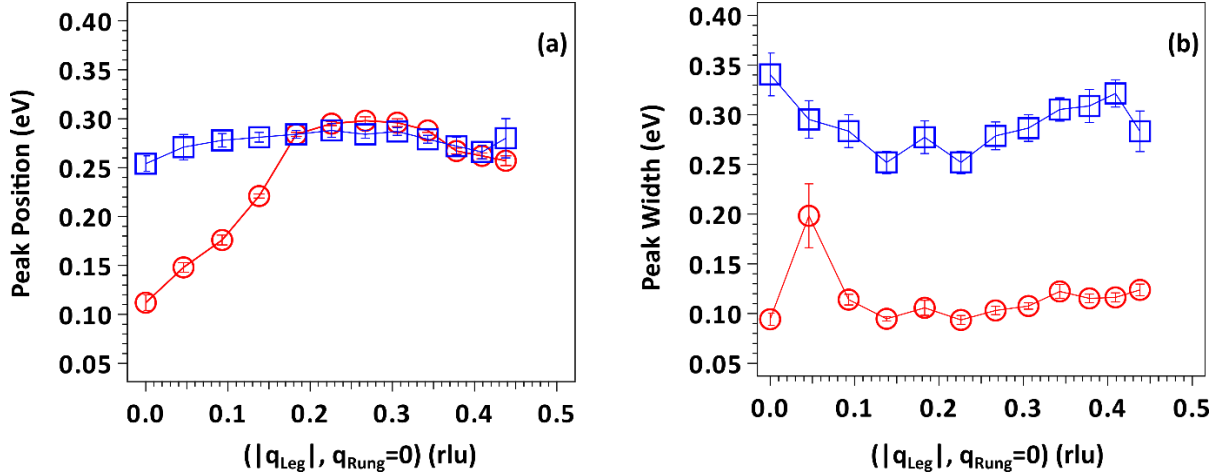

**FIG. S3:** Fitting results for the magnetic excitations in the Cu $L_3$-edge RIXS spectra of Sr14 (red circles) and Sr1.8Ca12.2 (blue squares). For the color-shaded Gaussian (Lorentzian) component of Sr14 (Sr1.8Ca12.2) shown in Fig. S2, the fitted peak positions and widths are plotted in (a) and (b), respectively.

In Fig. S2(a), we decompose the excitations of Sr14 into the elastic line around 0 eV loss and magnetic excitations in the range of $150 - 400$ meV. The elastic line has contributions from elastic scattering, including pure charge and diffuse scattering due to surface roughness, and a quasi-elastic weight accounting for the low-energy asymmetry of the elastic line. The latter is attributed to the unresolved low-energy excitations such as phonons or magnetic excitations in the chains [S1]. Both components are described by Gaussian functions with a width given by the overall instrumental energy resolution. The magnetic excitations are fitted with a resolution-limited Gaussian profile for the dominant sharp peak, and a broad Gaussian peak [250meV full-width to half-maximum (FWHM)] for the residual higher-energy weight that becomes more distinguishable with increasing momentum. Using these lineshapes, we managed to fit the sharp two-triplon (2T) peak position with errors in the peak positions no greater than 10 meV. The momentum-dependent magnetic spectral weight is in agreement with the peak energy and dispersion evaluated from the previous Cu $L_3$-edge RIXS work [S1]. The width of the 2T excitations increases towards the zone boundary. This increase may be caused by the enhanced scattering of the higher-energy component, with possible additional contributions from high-order multi-triplon scattering or charge excitations [S2, S3]. Another source for the high-energy spectral weight near zone boundary could be confined spinons, since the magnetic exchange couplings are close to the isotropic limit [S3]. Further investigations are required to clarify this point.

The same fitting strategy for the magnetic excitations fails for the heavily Ca-doped compound, as the increased Ca content leads to a broadening of the magnetic RIXS response. The damped lineshape for magnetic scattering of Sr1.8Ca12.2 also leads to increased uncertainties in the estimated peak positions. As shown in Fig. S2(b), we keep the two-component contributions to fit the elastic line and use a damped Lorentzian with a 350~400 meV FWHM for the magnetic excitation spectra. This choice gives us a fitting error of no greater than 30 meV for the peak positions of Sr1.8Ca12.2's magnetic excitations across momentum space.

**Polarization-Dependent O K-edge XAS & Determination of Ladder Hole Content**

Polarization-dependent O 1s XAS can be used to distinguish the local Cu-O bonding environments of the edge-sharing $CuO_2$ chains and two-leg $Cu_2O_3$ ladders in SCCO. Here, we apply the fitting model established in Ref. [S4] to estimate the hole densities in nearly hole-depleted Sr14 and moderately hole-doped Sr1.8Ca12.2, which will serve as a starting point for our model calculations. Before proceeding, we note that the fitting model of Ref. [S5] appears to overestimate the ladder hole concentrations [see Fig. S4(e)], resulting in values that are incompatible with the experimental observations of sharp collective magnetic excitations. We, therefore, do not apply this procedure here.

In Fig. S4(a)-(d), the intensity and peak position of the lower-energy pre-edge hole peak $H1$ is polarization-independent. This peak is interpreted as the doped holes in the chains, due to their 90° Cu-O-Cu bond angle [S4]. As a result, the linear dichroism is vanishing in the spectral contributions from the chains. On the other hand, the higher-energy component in the double-peak structure, labeled as $H2$, was argued to be mainly originating from the hole content in the ladders [S4]. $H2$ is enhanced when the polarization of incident X-rays $E_{\text{inc}}$ is parallel to the ladder-leg orientation (c-axis), in line with the increasing number of accessible O 2p orbitals in the ladder subsystem [S4]. The spectral weight at 529.25 eV represents the upper Hubbard band (UHB) [S4]. The linear dichroism in O 1s XAS is suppressed with Ca doping, as shown in Fig. S4(c)-(d) and in agreement with literature [S4–S7]. This suppression was speculated to originate from lattice distortions in the ladder subsystem, creating an increasing quasi-2D electronic character resembling the $CuO_4$ plaquette planes [S4] that may connect to the observed SC phases in SCCO [S8]. One issue worth noting is that this modeling might fail to describe the hole content across a wide chemical doping, as the approach has a phenomenological basis. As shown in Fig. S4(c) and Fig. S4(d), it is unclear whether the two-component assignment still reasonably captures the chain-ladder hole distribution for nearly complete replacement of Sr by Y, La, or Ca.

We compare our estimated hole contents with existing experimental reports (Ref. [S4-S7, S9-S14]) in Fig. S4(e). The number of ladder holes per chemical formula (pcf) $n_{\text{Ladder}}$ are estimated as $n_{\text{Ladder}} = 6\,(H2_a + H2_c)/(H1_a + H1_c + H2_a + H2_c)$ [S4], where the subscript labels $a$ and $c$ represent the $H1$ and $H2$ spectral weights measured with the polarization vector of incident X-rays parallel to the rung or leg directions, as shown in Fig. S4(a)-(d). The factor of 6 reflects the total number of holes pcf. Early studies concluded that the Sr-Ca substitution does not change the total numbers of holes, but rather transfers the holes from chains to ladders due to the altered electrical potential in SCCO [S9]. We extract 0.85 and 1.43 ladder holes for Sr14 and Sr1.8Ca12.2 pcf, respectively, which corresponds to an estimation of 6% and 11% ladder hole densities in Sr14 and Sr1.8Ca12.2, respectively.

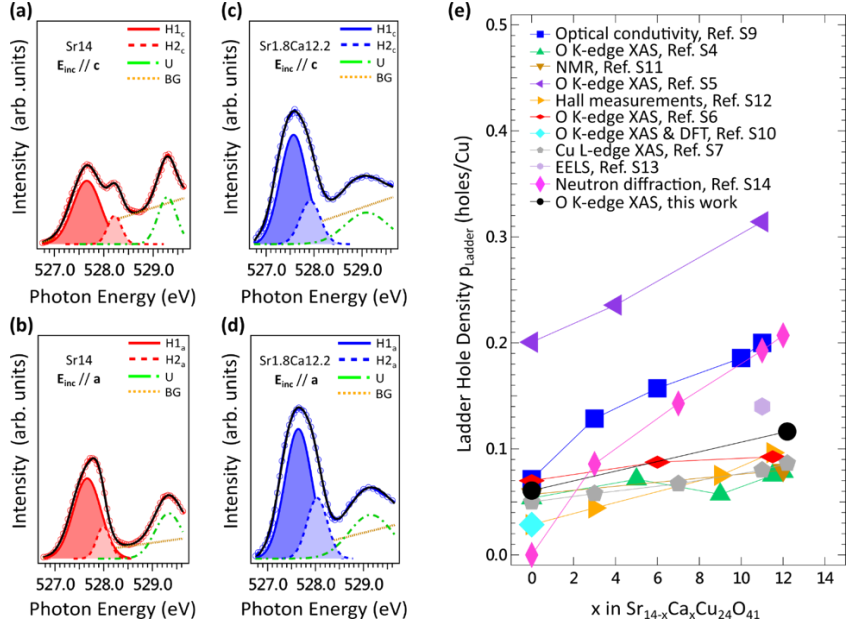

**FIG. S4:** O K-edge XAS spectra with fitting components for (a)-(b) Sr14 and (c)-(d) Sr1.8Ca12.2. The polarization of incident X-ray is parallel to the leg c-direction in (a) and (c), whereas (b) and (d) are measured in the rung a-direction configuration. (e) Experimentally determined number of holes in the ladder subsystem of SCCO, with comparison to the present literature [S4–S7,S9–S14].

Despite extensive studies, the experimental quantification of hole doping and carrier mobility with respect to the Ca content remains challenging. Early experiments on SCCO observed an increase in the low-energy Drude-like weight and enhanced metallicity in samples with elevated Ca content, which led to a speculated hole transfer between the chains and ladders upon isovalent Sr-Ca substitution [S9,S15-S16]. This transfer was rationalized in a ladder-dominant picture of the transport in SCCO due to the 180° Cu-O-Cu bond angle in the ladders [S17-S19]. Recent studies of $Sr_{14}Cu_{24}O_{41}$, however, revealed that errors up to $2-5$ can occur when quantifying the ladder hole content if charge and magnetic correlations in the chains are not taken into consideration [S10]. This fact has made it difficult to quantify the hole distribution in doped ladders. Further investigations will be needed to clarify these issues.

**Model Cluster Calculations for RIXS Spectra Using Density-Matrix Renormalization Group (DMRG) Method**

We use a single band Hubbard ladder with nearest neighbor hopping to model the system. The Hamiltonian is

$$H_0 = -t_{\text{leg}} \sum_{i,l,\sigma} (c^\dagger_{i,l,\sigma} c_{i+1,l,\sigma} + h.c) - t_{\text{rung}} \sum_{i,\sigma} (c^\dagger_{i,0,\sigma} c_{i,1,\sigma} + h.c) + U \sum_{i,l} \hat{n}_{i,l,\uparrow} \hat{n}_{i,l,\downarrow},$$

where $l = 0, 1$ indexes the legs of the ladder; $i$ indexes the sites along each leg; $c^\dagger_{i,l,\sigma}$ ($c_{i,l,\sigma}$) creates (annihilates) an electron of spin $\sigma$ on site $i$ of leg $l$; $\hat{n}_{i,l,\sigma} = c^\dagger_{i,l,\sigma} c_{i,l,\sigma}$ is the corresponding number operator; $t_{\text{leg}}$ ($t_{\text{rung}}$) are the hopping integrals along the leg (rung) direction; and $U$ is the strength of the *e-e* interaction.

Throughout this work, we parameterize the relative strength of the rung-leg couplings using the ratio $r = J_{\text{rung}}/J_{\text{leg}}$, where $J_{\text{leg}}$ ($J_{\text{rung}}$) is the superexchange integral along leg (rung) direction. $U$

can be mapped onto the leg and rung superexchange using the relationship $J_{\text{leg/rung}} \propto 4t_{\text{leg/rung}}^2/U$ such that $r = (t_{\text{rung}}/t_{\text{leg}})^2$. In all calculations, we fix $t_{\text{leg}} = 1$ and adjust $U$ and $t_{\text{rung}} = \sqrt{r}t_{\text{leg}}$ to obtain particular values of $J_{\text{leg}}/J_{\text{rung}}$.

The RIXS intensity within the Kramers-Heisenberg formalism can be expanded using the ultra-short core-hole (UCL) approximation [S20]. For the low-energy excitations at the Cu $L_3$-edge RIXS spectra for the layered cuprates, the non-spin-conserving $\Delta S = 1$ channel of RIXS has been shown to be captured at lowest order by the dynamical spin structure factor $S(q, \omega)$, while the spin-conserving $\Delta S = 0$ channel is approximated by a modified dynamical charge structure factor $\widetilde{N}(q, \omega)$ [S21]. This approximation has also recently been shown to be valid for the magnetic and charge excitations in spin ladders in Ref. [S3]. These structure factors are calculated as [S22]:

$$S(q, \omega) = \sum_f |\langle f | S^z_{q,\sigma} | g \rangle|^2 \delta(E_f - E_g + \omega)$$

and

$$\widetilde{N}(q, \omega) = \sum_f |\langle f | \tilde{n}_{q,\sigma} | g \rangle|^2 \delta(E_f - E_g + \omega),$$

where

$$\tilde{n}_{q,\sigma} = n_{q,\sigma} - n_{q,\sigma} n_{q,\sigma'}.$$

Here, $q$ and $\omega$ are the net (1D) momentum and energy transfer to the system, $\sigma$ is a spin index, $|g\rangle$ and $|f\rangle$ are the initial and final states of the scattering process, respectively, $E_g$ and $E_f$ are their respective energies, and $S^z_{q,\sigma}$ and $\tilde{n}_{q,\sigma}$ are the Fourier transforms of the local spin and charge operators, respectively.

We use the DMRG method to calculate these structure factors in real space and Fourier transform the results to momentum space following the procedure outlined in Ref. [S22]. Numerical simulations were performed using DMRG++ code [S23]. Computational details are described in the supplemental material of [S22].

For Sr14, we use a $32 \times 2$ ladder cluster and keep up to $m = 1000$ states and a maximum truncation error of $10^{-7}$. The artificial broadening parameter is set to $\eta = 47.5$ meV to match the experimental resolution. For Sr1.8Ca12.2, we use a $16 \times 2$ ladder with the same $m = 1000$ states and artificial broadening parameter $\eta$ as was used for Sr14. The maximum truncation error is reduced to $10^{-8}$.

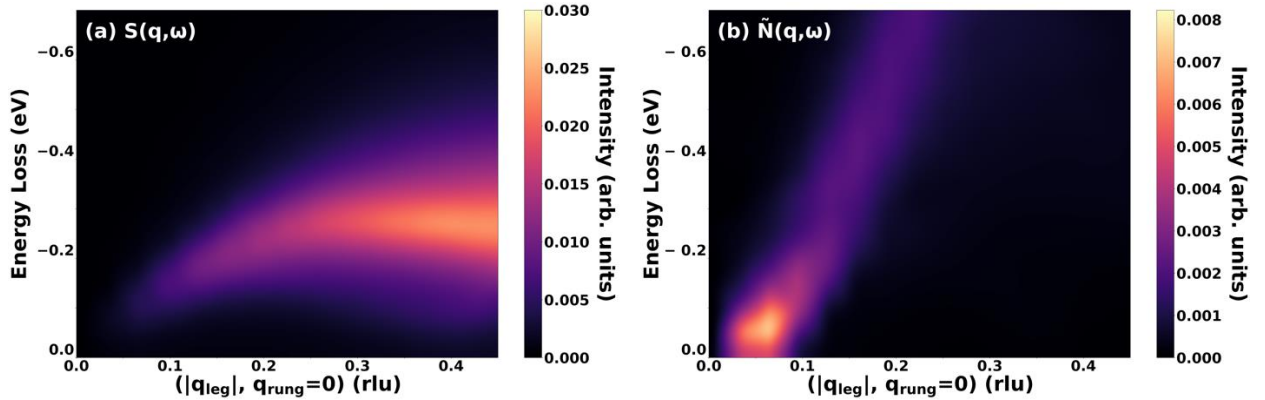

**FIG. S5:** (a) Dynamical spin and (b) modified dynamical charge structure factors for Sr14 representing the $\Delta S = 1$ and $\Delta S = 0$ channels of the RIXS spectra with bicubic interpolation. The spectra are calculated using DMRG on a $32 \times 2$ ladder with $t_{\text{leg}} = 340$ meV, $U = 8t_{\text{leg}}$, $r = 0.85$, $\eta = 0.1379 t_{\text{leg}}$ and hole doping of $p = 6.25\%$, close to the experimentally determined value from O K-edge XAS results.

We postulate that the increasing Ca content tends to localize the doped holes in the ladder subsystem, which is consistent with the reported semiconducting transport behavior [S19]. Previous studies also support the view that increased Ca substitution dopes the ladder subsystem by driving the holes from the chains to the ladders, with the edge-shared $CuO_2$ chains acting as charge reservoirs [S9,S15-S16]. On the other end, optical conductivity calculations on edge-shared cuprate chains have shown that electrons have low mobility [S24]. Therefore, the excess electrons introduced into the chain system by Ca substitution can provide an unscreen *attractive* potential acting on the holes in the ladder. One can view the charge transfer process as leading to an excitonic state involving a hole in the ladder and an electron in the chain.

To explore the implications of this idea on the low-energy spin excitations, we introduce additional impurity potentials $V_{\text{imp}} = 0.5U$ to four randomly chosen sites in the ladders, that will be referred to as the impurity sites. For each model calculation the sites of the ladder are labelled from 0 to 31, with even and odd number sites corresponding to the upper and lower legs, respectively. The individual configurations, chosen randomly, are then given by $(16, 17, 20, 28)$, $(6, 12, 13, 17)$, $(1, 12, 24, 31)$, $(7, 14, 15, 24)$, $(0, 1, 16, 21)$, $(0, 5, 20, 30)$, $(11, 16, 20, 22)$, and $(21, 28, 30, 31)$, where each number in a set corresponds to the site index of the impurity site. We make the additional assumption that the Coulomb potential at the impurity sites is poorly screened, so we extend its range to the first five nearest neighbors on the ladder rescaling its strength with distance following a $1/r$ dependence. For example, if $V_{\text{imp}} = 0.5U$ is placed on site 16 as is the case in set 1 then the three axial neighbors, two on either side of the same leg (14 and 18) and one along the same rung (15), have $V_{\text{imp}}/2 = 0.25U$ and $V_{\text{imp}}/2\sqrt{2} = 0.18U$ is placed on the two diagonal sites (13 and 17) on the opposite leg of the ladder. If a particular site on the ladder has overlapping contributions from multiple impurity sites, then they are summed. For example, site 16 in set 1, which is itself an impurity site and a diagonal neighbor to another (17). The total strength of the potential at that site is then $V_{\text{total}} = V_{\text{imp}} + V_{\text{imp}}/2\sqrt{2} = 0.68U$. The modified Hamiltonian then becomes

$$H = H_0 + \sum_{i,j,\sigma} n_{i,\sigma} V(i-j)$$

where $V = \delta_{i,j} V_{\text{imp}} + (1 - \delta_{i,j}) V_{\text{imp}}/2|i-j|$ where $|i-j| = 1$ for nearest neighbors and is cutoff beyond next-nearest-neighbors.

We then calculate both spin and modified charge structure factors for eight different disorder configurations as shown in Fig. S6 & S7. The final structure factors, shown in Fig. S8, are obtained by averaging the spectra produced by all disorder configurations.

DMRG results show that all sites that experience the effect of the impurity potential have reduced electron occupation in the ground state, i.e. they are predominantly hole-occupied. Electron number density at every site for each set is then plotted in Fig. S9 that shows the hole distribution across the ladder configurations. We observe that the occupation number is reduced on most impurity sites, but subject to the constraint that the total particle number is fixed.

The spin structure factors capture the major spectral features of the RIXS data for both Sr14 and Sr1.8Ca12.2 and are included in the main text. The modified dynamical charge structure factors are about a factor of 4 smaller than $S(q, \omega)$ in doped ladders (Fig. S5, S7 & S8). A special note is made on the modified charge dynamical response of set 8 (21,28,30,31), which appears negligible in comparison to other sets. This is due to the fact that this configuration places three impurity sites at the end of the ladder and pins the holes to the edge as seen for set 8 in Fig. S9. The rest of the ladder then behaves like an undoped system and one expects to only see charge excitations above the Mott-gap (~2.4 eV).

Our results show that, for localized impurities, the computed holes' configurations are such that it is less likely to cut the ladders into disconnected segments. This happens for two reasons: (1) the probability to have impurities on the same rung is much smaller than the probability of having them on different rungs (in our modeling, we have assumed a random distribution); (2) in gapped spin ladders, strong antiferromagnetic fluctuations survive (and are actually enhanced by) to small concentrations of non-magnetic impurities as the system "self-heals" the local disruptions caused by the impurites themselves [S25-S29]. Our spectra confirm this picture in that they exhibit a pronounced downturn at the $\Gamma$ point ($q_{leg} = 0$, $q_{rung} = 0$), which does not agree with the flat mode observed in the experimental RIXS data.

On the other hand, poorly screened impurities with a longer range Coulomb repulsion are more effective in cutting the ladders into disconnected segments, generating configurations where impurities can effectively clusterize depleting two or more neighbouring rungs. In this case, due to the "extended" disruption, the system is not able to self-heal by building stronger antiferromagnetic correlations. The resulting magnetic excitation spectrum is, therefore, flatter along the $q_{rung} = 0$ momentum transfer direction, in better agreement with the RIXS data.

Our results show that using four impurity sites and fixing the strength of the impurity potential to $V_{imp} = 0.5U$, is sufficient to obtain better agreement with the experimental RIXS data. However, these parameters are not independent and other combinations are likely possible. Determining the exact magnitude of impurity potentials would require a full quantum chemistry calculation, which is beyond the scope of this work. Moreover, some of the random impurity configurations spectra more closely resemble the experimental data. For example, set 1 (16,17,20,28) and set 2 (6,12,13,17) that has 3 impurity sites on adjacent rungs and act as a very large impurity sitting in the middle of the ladder interestingly provide the best agreement with the experiment for the spin dynamical structure factor. Electron density plots (Fig. S9) show that these particular sets have an extended hole occupied region in the middle of the ladder.

To visualize the effect of the impurities on the magnetic excitations, we plot the dynamical structure factor of the eight impurity sets with the spectral weight of the clean model subtracted, i.e. $S_{diff}(q, \omega + i\eta) = S_{imp}(q, \omega + i\eta) - S_{no-imp}(q, \omega + i\eta)$, in Fig. S10. The spectral weight is shifted to higher energy and flattens relative to the clean system, with the appearance of some low-energy spectral weight. The latter will be masked by the phonon excitations and quasi-elastic weight and may be difficult to observe experimentally.

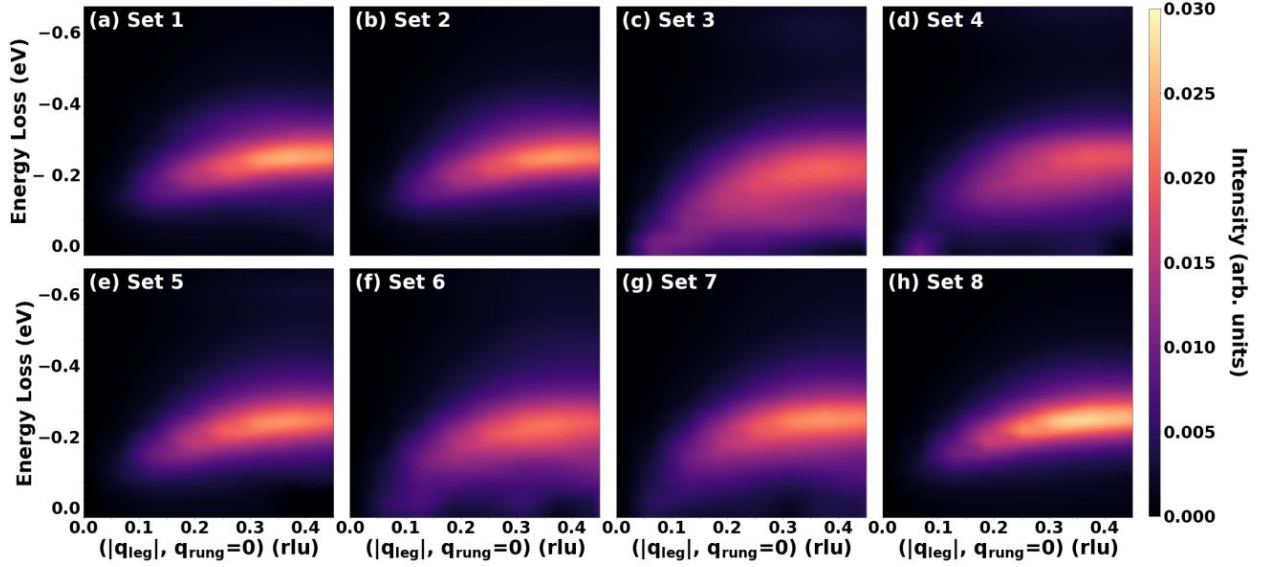

**FIG. S6:** (a)-(h) Dynamical spin structure factors for Sr1.8Ca12.2 representing the $\Delta S = 1$ channel of the RIXS spectra with bicubic interpolation. Calculated by DMRG on a $16 \times 2$ ladder with $t_{leg} = 300$ meV, $U = 9t_{leg}$, $r = 1.1457$, $\eta = 0.1583t_{leg}$, and hole doping $p = 12.5\%$ close to the experimentally determined value from O K-edge XAS results. (a) and (h) correspond to $(16, 17, 20, 28)$, $(6, 12, 13, 17)$, $(1, 12, 24, 31)$, $(7, 14, 15, 24)$, $(0, 1, 16, 21)$, $(0, 5, 20, 30)$, $(11, 16, 20, 22)$, and $(21, 28, 30, 31)$.

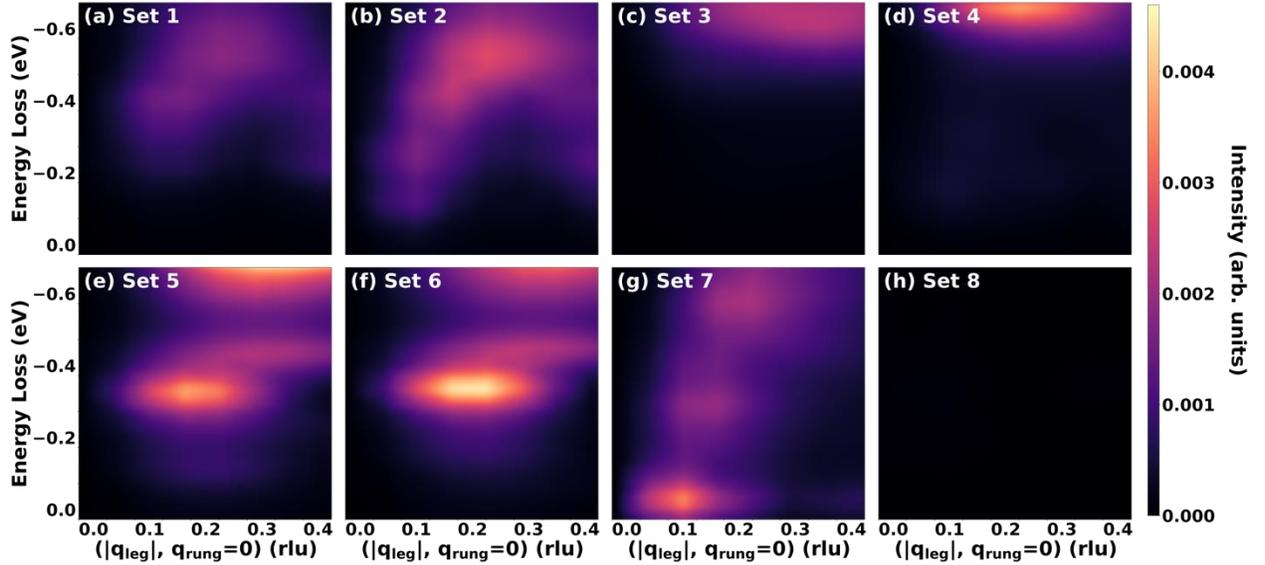

**FIG. S7:** (a)-(h) Dynamical modified charge structure factors for Sr1.8Ca12.2 representing the $\Delta S = 0$ channel of the RIXS spectra with bicubic interpolation. Calculated by DMRG on a $16 \times 2$ ladder with $t_{leg} = 300$ meV, $U = 9t_{leg}$, $r = 1.1457$, $\eta = 0.1583t_{leg}$, and hole doping $p = 12.5\%$ close to the experimentally determined value from O K-edge XAS results. (a) and (h) correspond to $(16, 17, 20, 28)$, $(6, 12, 13, 17)$, $(1, 12, 24, 31)$, $(7, 14, 15, 24)$, $(0, 1, 16, 21)$, $(0, 5, 20, 30)$, $(11, 16, 20, 22)$, and $(21, 28, 30, 31)$.

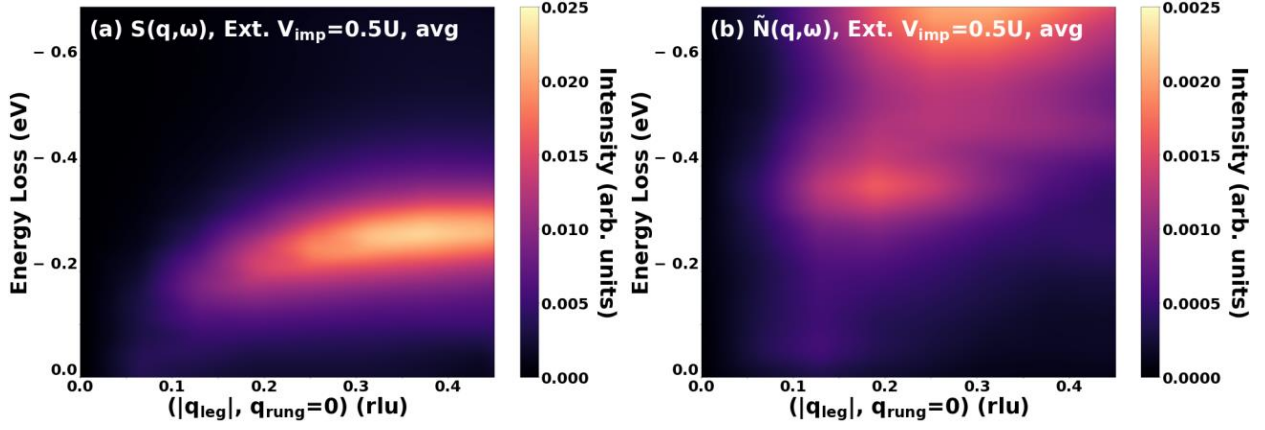

**FIG. S8:** The averaged $S(q, \omega + i\eta)$ and $\tilde{N}(q, \omega + i\eta)$ of the eight sets are shown in panel (a) and (b), respectively, representing the $\Delta S = 1$ and $\Delta S = 0$ channels of the RIXS spectra with bicubic interpolation. Calculated by DMRG on a $16 \times 2$ ladder with $t_{\text{leg}} = 300$ meV, $U = 9t_{\text{leg}}$, $r = 1.1457$, $\eta = 0.1583 t_{\text{leg}}$, and hole doping $p = 12.5\%$.

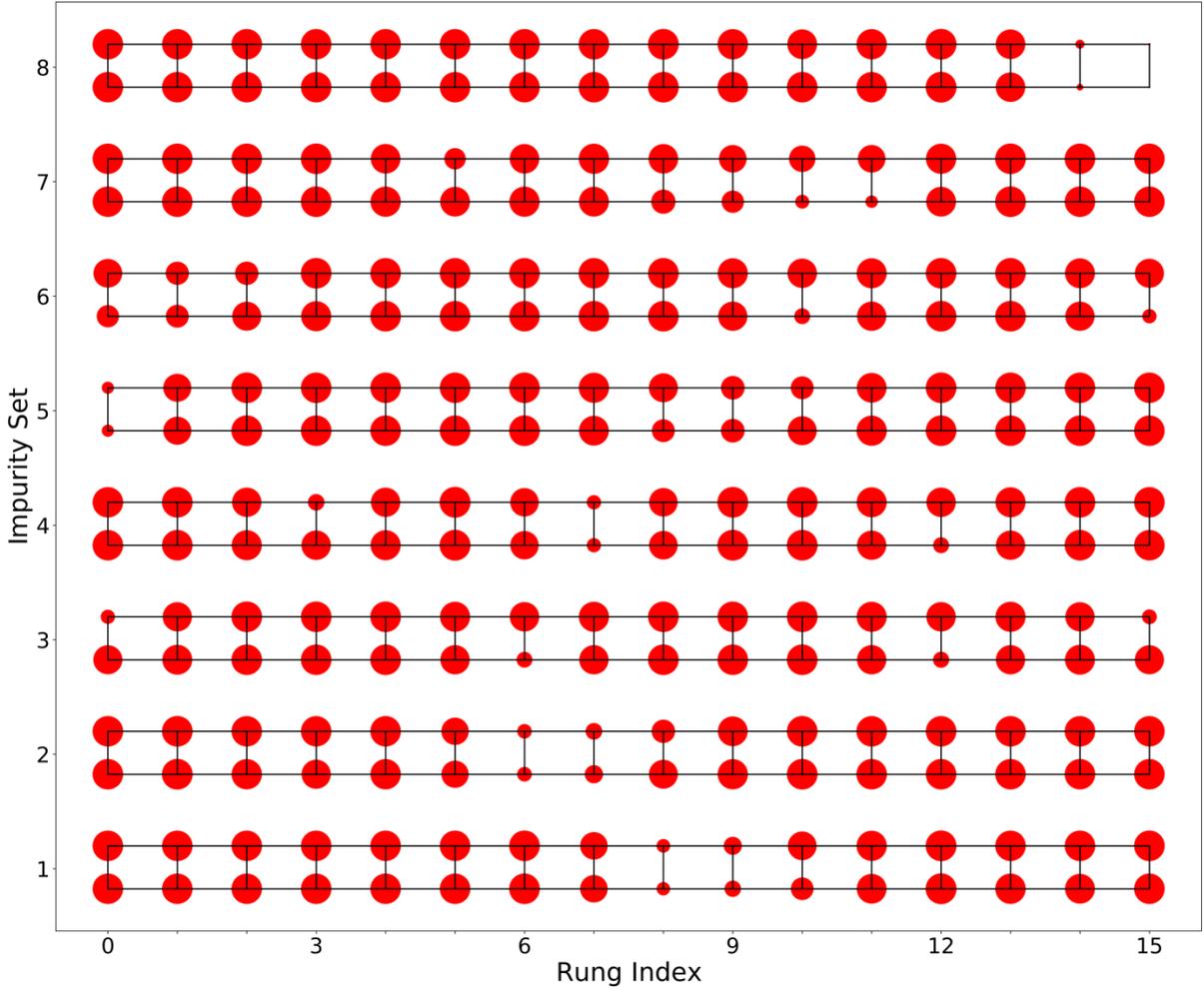

**FIG. S9:** Electron density $\hat{n}$ at every site for set 1 to 8 corresponding to $(16, 17, 20, 28)$, $(6, 12, 13, 17)$, $(1, 12, 24, 31)$, $(7, 14, 15, 24)$, $(0, 1, 16, 21)$, $(0, 5, 20, 30)$, $(11, 16, 20, 22)$, and $(21, 28, 30, 31)$, respectively. The area of the data points reflects to the magnitude of $\hat{n}$.

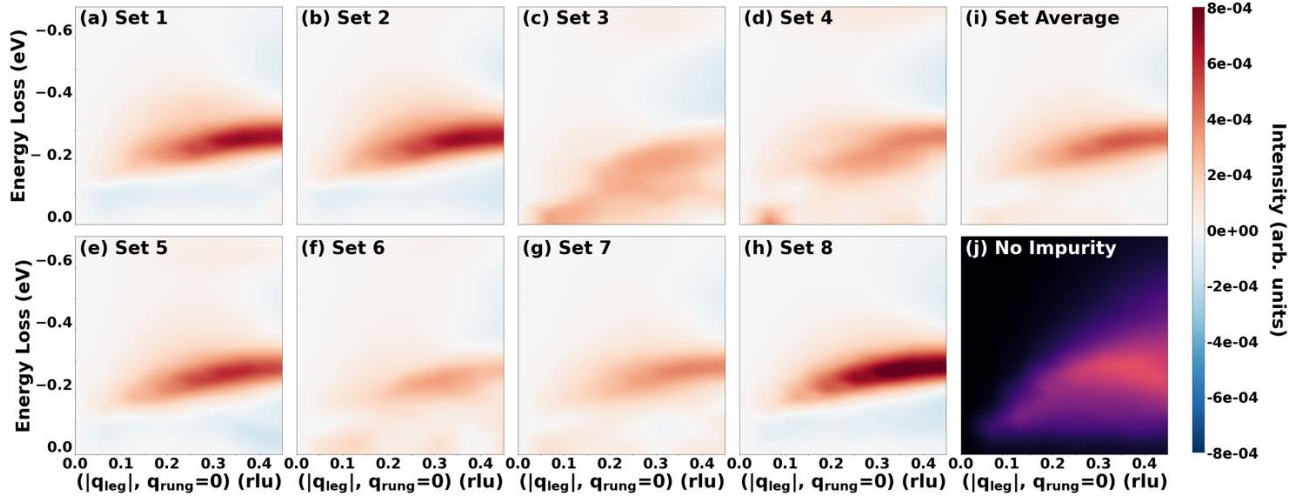

**FIG. S10:** Plots of the difference of the dynamical spin structure factors $S_{diff}(q, \omega + i\eta)$. (a)-(h) are the $S_{diff}(q, \omega + i\eta)$ for eight impurity sets of Sr1.8Ca12.2, (i) is the average $S_{diff}(q, \omega + i\eta)$ and, (j) is the full $S_{no-imp}(q, \omega + i\eta)$. DMRG results used here are for a $16 \times 2$ ladder with $t_{leg} = 300$ meV, $U = 9t_{leg}$, $r = 1.1457$, $\eta = 0.1583 t_{leg}$, and hole doping $p = 12.5\%$.

The effect of impurity potentials in inducing hole localization has been studied using DMRG on 1D and 3-leg Hubbard ladders before [S29]. To the best of our knowledge, however, our calculation is the first to include these effects in the charge and spin dynamics of such systems.

To understand the implications of hole localization in the pairing tendencies exhibited by the carriers in the ladder, we further calculate the averaged rung singlet pair correlation function $P(d)$ as a function of distance $d$ [S30]

$$P(d) = \frac{1}{L-d} \sum_{j=1}^{L-d} \langle \Delta_j^\dagger \Delta_{j+d} \rangle$$

where $\Delta_i^\dagger$ is defined as $\frac{1}{\sqrt{2}}(c_{i,0,\uparrow}^\dagger c_{i,1,\downarrow}^\dagger - c_{i,0,\downarrow} c_{i,1,\uparrow})$. The absolute value of the results are plotted in Fig. S11 and show that impurity potentials diminish the rung singlet pairing compared to the model for a clean doped ladder. Our findings are consistent with existing literature as Sr1.8Ca12.2 is expected to undergo an insulator to superconductor transition only at elevated pressures. To simulate this effect of hydrostatic pressure, we compute the rung singlet pair correlation averaged over all disorder configurations with the hopping integrals uniformly increased by 10%, 20%, and 30%. With increased hopping, we see an enhancement of the pair correlations suggesting that the ladder has strong pairing tendencies at large pressures.

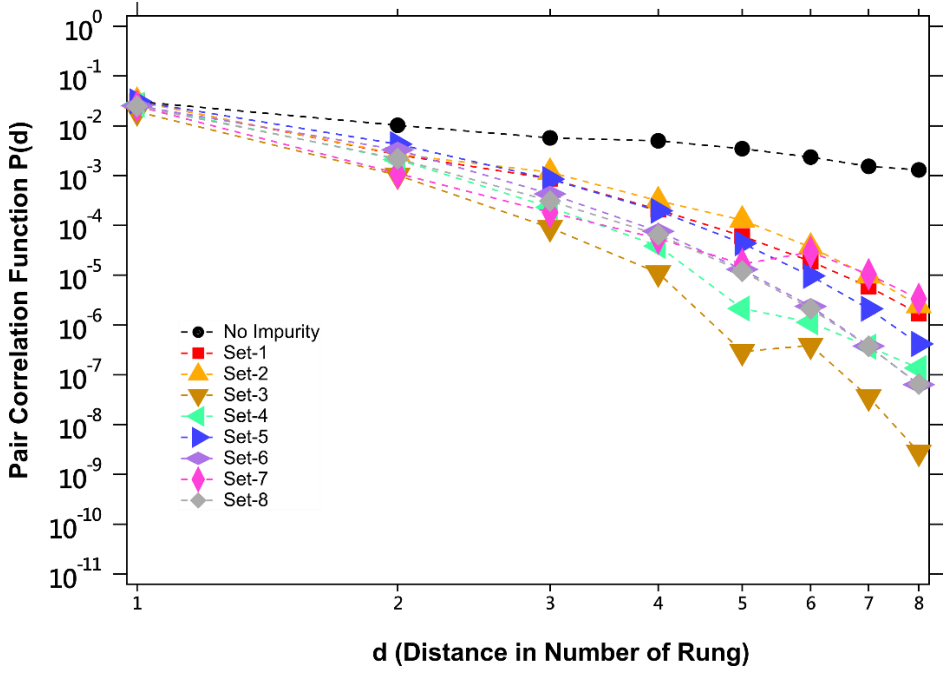

**FIG. S11:** Rung singlet pair correlations as a function of distance along the ladder leg for each of the eight impurity sets and the no-impurity case. Calculated by DMRG on the $16 \times 2$ ladder with $t_{\text{leg}} = 300$ meV, $U = 9t_{\text{leg}}$, $r = 1.1457$, $\eta = 0.1583 t_{\text{leg}}$, and hole doping $p = 12.5\%$.

In addition to hole localization, we consider two alternative possibilities that could theoretically capture the experimental features in Sr1.8Ca12.2. The first is increased rung coupling in Sr1.8Ca12.2. The Sr14 data are well described by adopting a rung ratio $r = 0.85$, consistent with prior estimates [S3]. The flattened dispersion of the excitation spectrum of Sr1.8Ca12.2 might be explained by adopting a larger rung ratio $r$, due to the formation of localized spin dimers along the rungs [S3]; however, we have found that such a large ratio is inconsistent with the available crystal structure data, as shown in Fig. S12. Within perturbation theory, the superexchange between pairs of Cu ions is given by [S31]

$$J = \frac{4 t_{pd}^4}{\Delta^2} \left( \frac{1}{\Delta} + \frac{1}{U_{dd}} \right) \propto t_{pd}^4,$$

where $\Delta$ is the charge transfer energy, $U_{dd}$ is the on-site Coulomb repulsion on the Cu site, and $t_{pd}$ is the Cu-O hopping integral. In the Ca-doped system, we assumed that the Cu-O hopping integral depends on the Cu-O bond distance as $t_{pd} = t_{pd}^0 (1 + \Delta d_{Cu-O}/d_{Cu-O})^{-3}$, where $t_{pd}^0$ and $d_{Cu-O}$ are the hopping and bond-distance in the undoped system and $\Delta d_{Cu-O}$ is the change in Cu-O distance in the Ca-doped system with respect to Sr14. The rung and the ladder Cu-Cu distances for Sr14 and the various Ca-doped systems were obtained from Refs. [S14] and [S32-S33] and we took the Cu-O distance to be half these values. Using this structural data, we then estimated $J_{\text{rung}}$ and $J_{\text{leg}}$ in the Ca-doped system and obtained $r = J_{\text{rung}}/J_{\text{leg}} < 1.15$ for all Ca concentrations. These estimates are inconsistent with large $r \approx 2 - 4$ that would be needed to drive the system into the spin dimer regime. We adopted $r = 1.15$ as a conservative estimate for Sr1.8Ca12.2 for the results shown here and in the main text.

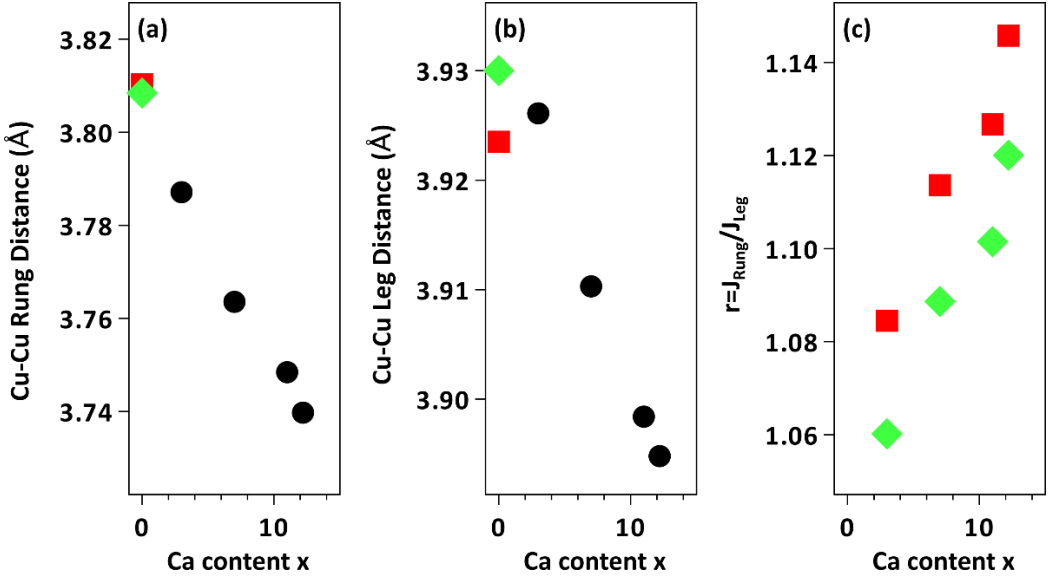

**FIG. S12:** (a) and (b) shows the rung and ladder (leg) distances for various Ca doping values obtained from different sources (red squares: Ref. S32; green diamonds: Ref. S33; black circles: Ref. S14). (c) shows the rung ratio estimates for Sr1.8Ca12.2, obtained using the ladder rung distances for Sr14 given in Refs. [S32] and [S33].

A second scenario is suggested by two studies that examined the effect of Ca substitution on the spin excitations in spin-chain cuprates. The first examined the zig-zag spin-½ chain cuprate $SrCuO_2$ and $Sr_{0.9}Ca_{0.1}CuO_2$ [S34] and the second examined the spin-½ chain cuprate $Sr_2CuO_3$ and $Sr_{1.9}Ca_{0.1}CuO_3$ [S35]. In the former, the observed exponential decay of the Copper (Cu) nuclear magnetic resonance (NMR) spin lattice relaxation rate confirms the opening of a spin gap in its Ca-doped analog. In the latter, the results suggest the presence of a spin gap with the possibility of some magnetic ordering interspersed in between. Using DMRG calculations on the Heisenberg model, the spin gap was then obtained by introducing an impurity driven local variation of the intrachain coupling $J$ [S34]. While the zig-zag spin-½ chain $Sr_{0.9}Ca_{0.1}CuO_2$ also contains an additional interchain coupling $J'$, it's estimated to be at least an order of magnitude smaller [S36] and its modulation required to obtain the spin gap was deemed unphysical [S34]. The appearance of the spin gap with the same order of magnitude in both the zig-zag and the single spin-½ chains, the latter of which contains only intrachain couplings, is also used to justify the negligible role played by $J'$ [S35].

Adopting a similar modulation of superexchange parameters to our ladder subsystem comes with a few caveats. Firstly, a spin-gap is observed in both Sr14 and Sr1.8Ca12.2 while in the spin ½ chains the Ca substitution drives the opening of a spin-gap. Secondly, and more importantly, there is a difference in the local geometries. Unlike the spin ½ chains where the intrachain coupling $J$ is the only modulated interaction, in ladders, both $J_{rung}$ and $J_{leg}$ are of comparable strength and can be modulated. Our rung-ratio analysis also limits the range of possible values for the two couplings as $0.85 < r = J_{rung}/J_{leg} < 1.15$. To test the viablilty of superexchange modulations we compute the dynamical spin structure factor for the t-J model for four different configurations where local $J_{\text{rung}}$ and $J_{\text{leg}}$ are allowed to vary between $0.4t_{\text{leg}}$ and $0.5t_{\text{leg}}$ such that the local rung coupling varies between 0.8 and 1.25. The results, along with their average, are shown in Fig. S13. We observe that the computed spectra are very similar to the non-impurity Hubbard Model calculation as it is expected for the large $U$ limit. We, therefore, conclude that the local modulation of superexchange integrals cannot capture the excitations of Sr1.8Ca12.2.

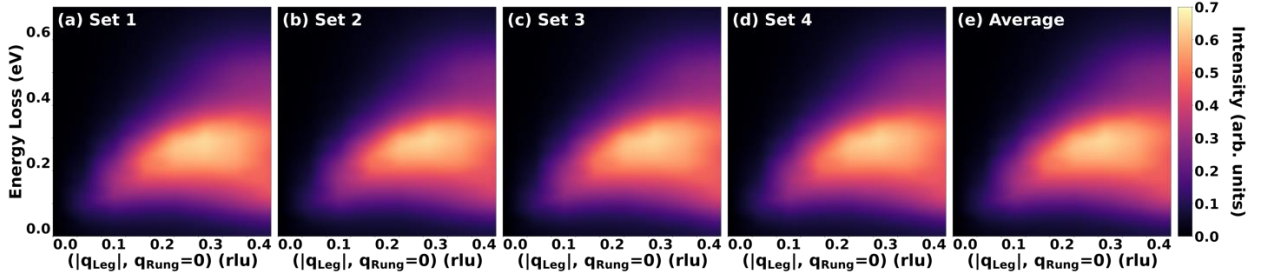

**FIG. S13:** (a)-(h) Dynamical spin structure factors for Sr1.8Ca12.2 calculated using locally modulated $J_{\text{leg}}$ and $J_{\text{rung}}$ for the t-J model with bicubic interpolation. Calculated by DMRG on a $16 \times 2$ ladder with $t_{\text{leg}} = 300$ meV, $0.4t_{\text{leg}} < J_{\text{leg}}, J_{\text{rung}} < 0.5t_{\text{leg}}$, $0.8 < r < 1.25$, $\eta = 0.1583 t_{\text{leg}}$, and hole doping $p = 12.5\%$.

**Disentanglement of Non-Spin-Conserving $\Delta S = 1$ and Spin-Conserving $\Delta S = 0$ Scattering Channels in Polarization-Dependent Cu $L_3$-edge RIXS**

To decompose the 2T excitations in the $\Delta S = 1$ and $\Delta S = 0$ channels [S2], we employ the polarization-dependent Cu $L_3$-edge RIXS measurements introduced in Ref. [S37]. Specifically, we measure the RIXS spectra with incident X-rays in $\sigma$ and $\pi$ polarization in the same scattering geometry, which yields the polarization-dependent magnetic RIXS intensity $I(\varepsilon, q)$ of 2T excitations. For practical purposes, we follow Ref. [S37] and perform the $\Delta S = 1$ and $\Delta S = 0$ scattering in a grazing emission geometry $(q_{leg} > 0)$. This setup reduces the errors from intensity fluctuations or unphysical results due to the small differences between $I(\sigma, q < 0)$ and $I(\pi, q < 0)$ that occur in the grazing incidence geometry. Additionally, the spectral differences between $\sigma$ and $\pi$ polarization from self-absorption effects are also reduced in grazing emission geometry [S37].

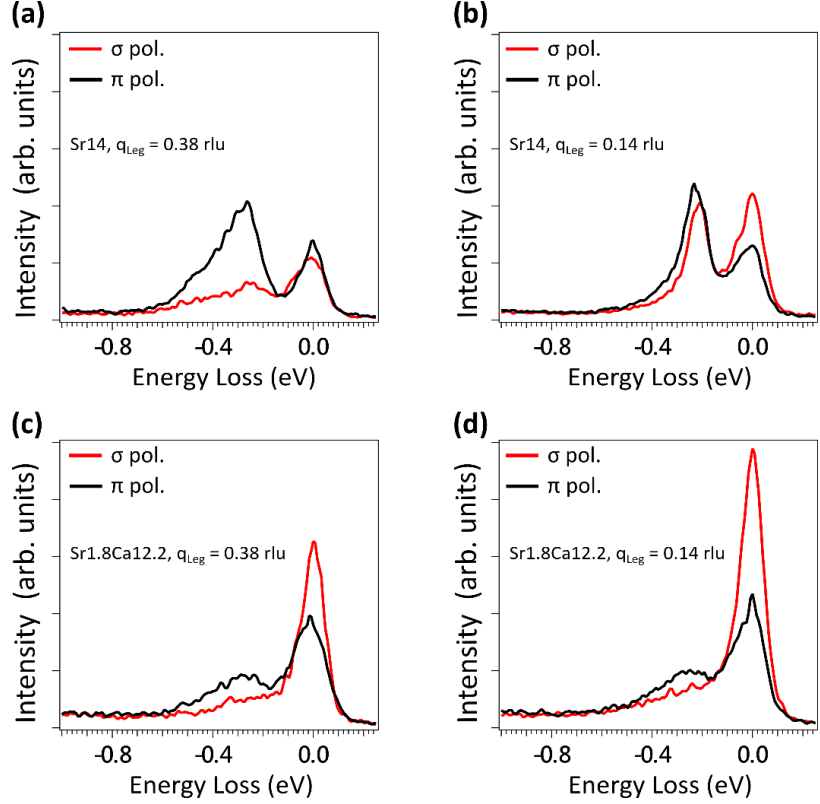

**FIG. S14:** (a)-(d) Polarization-dependent Cu $L_3$-edge RIXS spectra zoomed in around the low-energy excitations. Here the experimental data are normalized to the total spectral weight of dd excitations with respect to the theoretical cross sections.

The RIXS intensities can be factored as $I(\varepsilon, q) = G(\Delta S, \varepsilon, q, \omega) F(\varepsilon, q)$. Here, $\omega$, $\varepsilon$ ($\sigma$ or $\pi$), and $q$ represent the energy loss, polarization of incident X-rays, and momentum-transfer, respectively. With the form factors $F(\varepsilon, q)$ for respective $\Delta S = 1$ and $\Delta S = 0$ scattering, one can obtain the dynamical structure factors $G(\Delta S, \varepsilon, q, \omega)$ from $I(\varepsilon, q)$. The form factors $F(\varepsilon, q)$ here are approximated by the local spin-flip and charge scattering probabilities (represented by the elastic scattering probabilities) $P(\varepsilon, q)$, which can be calculated in the single-ion approximation [S38]. The atomic symmetries of core-hole p and valence d orbitals for the $Cu^{2+}$ ion are taken into account in these calculations. For the given RIXS response probed at $3d^9$ white line, this agrees with the majority of $Cu^{2+}$ electronic contributions from the underdoped ladder subsystem ($n_{\text{ladder}}$ of $5 - 15\%$ and $n_{\text{chain}}$ of $40 - 50\%$, see previous section). Here, we normalize the RIXS spectra with respect to the theoretical dd cross section before evaluating $G(\Delta S, \varepsilon, q, \omega)$. This follows our data treatment of the experimental RIXS results (normalized to dd excitations from $-1.5$ to $-3$ eV loss) [S37]. Thus, we can extract the $\Delta S = 1$ and $\Delta S = 0$ channels by solving the equations of the polarization-dependent RIXS intensity. We estimated the error bars for the extracted spectra by assuming Poisson statistics [S37]. The raw RIXS spectra measured for selected momenta and varied polarizations are shown in Fig. S14(a)-(d). Here, the elastic line has been removed from the RIXS spectra before $G(\Delta S, \varepsilon, q, \omega)$ was extracted from the RIXS intensities. The resulting disentangled spin-resolved channels in the main text are shown in the energy regime above the elastic line to exclude possible contributions from the quasi-elastic scattering.